\documentclass[aps,prd,reprint,groupedaddress]{revtex4-2}

\usepackage{amsmath}
\usepackage{amsfonts}
\usepackage{mathrsfs}
\usepackage{amssymb}

\newcommand{\normord}[1]{\;\bm{\vcentcolon}\,\mathrel{#1}\,\bm{\vcentcolon}\;}
\providecommand{\vcentcolon}{\mathrel{\mathop{:}}} 

\newcommand{\normleft}[1]{\;\bm{\vcentcolon}\,\mathrel{#1}}
\newcommand{\normright}[1]{\mathrel{#1}\,\bm{\vcentcolon}\;}

\newcommand{\sgn}{\text{sgn}}

\usepackage{slashed}
\usepackage{simpler-wick}
\usepackage{tensor}
\usepackage{bm}

\begin{document}


\title{Quantum electrodynamics in the null-plane causal perturbation theory}


\author{O.A. Acevedo}
\email[]{oscar.acevedo@unesp.br}

\author{B.M. Pimentel}
\email[]{bruto.max@unesp.br}
\affiliation{Institute for Theoretical Physics (IFT), S{\~a}o Paulo State University (UNESP), R. Dr. Bento Teobaldo Ferraz 271, S{\~a}o Paulo, SP 01140-070, Brazil}


\date{\today}

\begin{abstract}
We study quantum electrodynamics (QED) in the light-front dynamical form by using null-plane causal perturbation theory.  We establish the equivalence with instant dynamics for the scattering processes, whose normalization allows to construct the instantaneous terms of the usual null-plane QED Lagrangian density. Then we study vacuum polarization and normalize it by studying its insertions into M{\o}ller's scattering process, obtaining the complete photon's propagator, which turns to be equivalent to the one of instant dynamics only when gauge invariance is taken into account.
\end{abstract}


\maketitle


\section{Introduction}

Quantum electrodynamics (QED) is the theory of the interaction between leptons and photons. It was the first quantum field theory to be constructed and, by virtue of its accessibility to experimental testing as well as its great success in predicting physical quantities such as the gyromagnetic ratio of the electron and Lamb's shift, it becomes the paradigm of this entire area of physics and one of the most studied theories. The most famous formulation of QED is the one developed by Tomonaga \cite{Tomonaga1,Tomonaga2,Tomonaga3,Tomonaga4,Tomonaga5}, Schwinger \cite{Schwinger1,Schwinger2,Schwinger3}, Feynman \cite{Feynman} and Dyson \cite{Dyson1} between 1946 and 1949. It was also Dyson, in 1949, who invented the techniques of regularization of the QED integrals and re-normalization of the theory by the absorption of the infinities into the mass and charge terms \cite{Dyson2}.

In 1949, Dirac \cite{Dirac} started the study of the relativistic dynamical forms. He discovered three possibilities: (a) \textit{Instant dynamics:} The one in which the isochronic surfaces are the planes of constant $x^0$; (b) \textit{point-form dynamics:} The isochronic surface is the superior branch of the hyperboloid $a^2=x^2$, the parameter $a^2$ being the time; (c) \textit{light-front dynamics:} The isochronic surfaces are null planes of constant $x^+$. This list was completed by Leutwyler and Stern \cite{LeutwylerStern} in 1978; they encountered two more dynamical forms \footnote{The five dynamical forms are shown to be the only possibilities, and no other can exist, under the imposition of transitivity of the stability group of the isochronic surfaces, this is to say, that every point of the surface can be mapped into any other point of it by some element of the stability group.}, with the following isochronic surfaces: (d) the superior branch of the hyperboloid $(x^0)^2-(x^1)^2-(x^2)^2=a^2$, and (e) the superior branch of the hyperboloid $(x^0)^2-(x^3)^2=a^2$; in both cases the parameter $a^2$ being the time. Among these dynamical forms the light-front one is special: in it the number of Poincar{\'e} generators independent of the interaction is maximum \cite{Dirac}; also, the null planes are the characteristic surfaces of Klein-Gordon-Fock's equation \cite{Rohrlich1, Rohrlich2}. These theoretical advantages of light-front dynamics translate into its success in treating a variety of practical problems, for example, in the context of current algebra \cite{FubiniForlan,Weinberg,BebieLeutwyler}, in the study of laser fields \cite{Rohrlich2, Neville1,Neville2}, for treating deep-inelastic scattering \cite{Drell1,Drell2,Drell3,Drell4} or in QCD for the study of hadron physics \cite{Igreja}.

However, being, as there are, many ways to describe the relativistic dynamics, there will also be various quantum field theories. The obvious question to ask in this context is which one of them is the correct theory, or if they are physically equivalent. To develop such theories is necessary in order to answer that question. Focusing into light-front dynamics, the quantization of fields on the null plane and the corresponding formulation of null-plane QED were done by Chang and Ma \cite{ChangMa}, Kogut and Soper \cite{KogutSoper}, Rohrlich and Neville \cite{Rohrlich2, Rohrlich3}, and Leutwyler, Klauder and Streit \cite{Leutwyler}. The equivalence between null-plane QED and the conventional instant one was then considered by Ten Eyck and Rohrlich \cite{TenEyck1,TenEyck2} and by Yan \cite{ChangYan1,ChangYan2}. In all these calculations, Feynman's amplitudes at one-loop level exhibited double-pole singularities as a consequence of an inconsistent treatment of the poles of the gauge field propagator in the null-plane gauge $A^+=0$; this problem was solved by Pimentel and Suzuki \cite{PimentelSuzuki1,PimentelSuzuki2}, who proposed a prescription to treat those poles in a causal way. Perturbative renormalization of null-plane QED was studied by Brodsky, Roskies and Suaya \cite{BrodskyRS} and by Mustaki, Pinsky, Shigemitsu and Wilson \cite{Mustaki}. Additionally, the constraint structure of classical QED and scalar QED in light-front dynamics and in the null-plane gauge was studied by Casana, Pimentel and Zambrano \cite{CPZ}. However, the equivalence between null-plane QED and the instant one is still under discussion; particularly, the importance of the instantaneous terms in the fermion and gauge fields propagators is not clear; recent reviews on the \textit{status quo} of the gauge field propagator can be found in Refs. \cite{SuzukiSales,Mantovani}. Very recently, the equivalence problem for one-loop radiative corrections was studied in Ref. \cite{BhamreMisra1,BhamreMisra2}, and the fulfilment of Ward-Takahashi's identity at one-loop order in Ref. \cite{SuzukiJi}.

It is in this context that the axiomatic approaches could offer a new insight into the subtleties of null-plane QED. Particularly, we will adopt the ``$S$-matrix program'' point of view, initiated by Heisenberg \cite{HeisenbergS} in 1943 as an attempt to go beyond the Lagrangian theory, and which was axiomatized in instant dynamics in the works by St{\"u}ckelberg and Rivier \cite{StueckelbergBook, Stueckelberg} and Bogoliubov, Medvedev and Polivanov \cite{BogoMedPoli, Bogo, BogoLogunov}. The detailed perturbative solution to Bogoliubov-Madvedev-Polivanov's axioms was carried out in 1973 by Epstein and Glaser \cite{EpsteinGlaser}, in a method in which the causality axiom plays an essential \textit{r{\^o}le}, and its first application to QED was done by Scharf in 1989 --in a monograph which is the first edition of Ref. \cite{ScharfFQED}--. This theory is now called ``causal perturbation theory''. Within this framework, QED in 2+1 dimensions was consider by Scharf, Wreszinski, Pimentel and Tomazelli \cite{Pimentel1}, while D{\"u}tsch, Krahe and Scharf \cite{ScharfSQED} used this theory to study scalar QED (SQED), showing that in the causal approach it suffices to start with the first-order coupling and the second-order vertex of the usual formulation is automatically generated as a normalization term in the second-order step. SQED was also investigated by Lunardi, Pimentel, Valverde, Manzoni, Beltr{\'a}n and Soto \cite{Pimentel31,Pimentel32}, who have used CPT to study the equivalence between Klein-Gordon-Fock's and Duffin-Kemmer-Petiau's scalar quantum electrodynamics. Also Podolsky's second order electrodynamics was considered from the causal point of view by Bufalo, Pimentel and Soto \cite{Pimentel41,Pimentel42}. So, QED in instant dynamics CPT is well established. It is the purpose of this paper, which is the first in a series, to start the study of null-plane QED in the causal framework. The formulation of CPT on light-front dynamics was done in Refs. \cite{APS21,APS22}, in which the causality axiom is referred to the null-plane time coordinate $x^+$, and has been successfully applied to obtain the radiative corrections for Yukawa's model \cite{AP2}, directly showing the equivalence with the instant dynamics formulation \cite{ABPS} for that model. In view of the mentioned successes of CPT, we hope that this framework would lead to a very clear formulation of null-plane QED.

This paper is organized as follows. Sec. \ref{sec:fields} is devoted to the definition of the field operators of the electron and photon. In Sec. \ref{sec:1} we offer a short review of null-plane CPT. The construction of the second order causal distribution for null-plane QED is done in Sec. \ref{sec:2}. M{\o}ller's and Compton's scattering processes are studied in Sec. \ref{sec:3}, in which also a comparison with the Lagrangian approach is discussed. Then, in Sec. \ref{sec:vacpol} we turn to vacuum polarization. Sec. \ref{sec:Conc} contains our conclusions and perspectives of future work.

\section{Quantized field operators of null-plane QED}\label{sec:fields}

QED deals with fermion and photon fields; the quantization of them was done in Ap. A of Ref. \cite{APS22} by the method of direct construction of Fock's space and a careful choice of the basis functions in the one-particle Hilbert space, exploiting the relation with the classical Goursat's problem. Then it was obtained that the fermion quantized field operator has the following expression:
\begin{align}\label{eq:2.3.11}
\psi(x)=&(2\pi)^{-3/2}\sum_s\int d\mu(\bm p)\sqrt{2p_-}\big(u_s(\bm p)b_s(\bm p)e^{-ipx}\nonumber\\
&+v_s(\bm p)d_s^\dagger(\bm p)e^{ipx}\big)\ ,
\end{align}
with the four-component functions $u, \overline u$ and $v,\overline v$ normalized so as to satisfy the sum rules:
\begin{equation}\label{eq:2.3.13}
\sum_s u_s(\bm p)\overline{u}_s(\bm p)=\frac{E\gamma^++|p_-|\gamma^-+p_\perp\gamma^\perp+m}{|2p_-|}\ ,
\end{equation}
\begin{equation}\label{eq:2.3.14}
\sum_s v_s(\bm p)\overline v_s(\bm p)=\frac{E\gamma^++|p_-|\gamma^-+p_\perp\gamma^\perp-m}{|2p_-|}\ ,
\end{equation}
and the emission and absorption field operators satisfying the following anti-commutation relations:
\begin{equation}\label{eq:2.3.15}
\left\{b_s(\bm p);b_r^\dagger(\bm q)\right\}=2p_-\delta_{sr}\delta(\bm p-\bm q)=\left\{d_s(\bm p);d_r^\dagger(\bm q)\right\}\ .
\end{equation}
The anti-commutator of Dirac's field with its Dirac's adjoint is:
\begin{equation}\label{eq:2.3.18}
\left\lbrace\psi(x);\overline\psi(y)\right\rbrace=-iS(x-y)\ ,
\end{equation}
with the distribution $S(x)$ the one which in the classical case solves Goursat's problem for Dirac's equation:
\begin{align}\label{eq:2.3.19}
S(x)&=i(2\pi)^{-3}\int d^4p(\slashed p+m)\text{sgn}(p_-)\delta(p^2-m^2)e^{-ipx}\nonumber\\
&=(i\slashed\partial+m)D(x)\ .
\end{align}
Particularly, the equal-time anti-commutation relation is:
\begin{equation}\label{eq:2.3.20}
\left\lbrace\psi(x^+;\bm x);\overline\psi(x^+;\bm y)\right\rbrace=(\slashed\partial-im)\left.D(x-y)\right|_{y^+=x^+}\ .
\end{equation}
In the right-hand-side of this equation, the derivation with respect to the variables $x^-$ and $x^\perp$ can be directly done, with result:
\begin{align}\label{eq:2.3.21}
D(0;\bm x-\bm y)&=(2\pi)^{-1}\delta(x^\perp-y^\perp)\int\limits_0^{+\infty}dp_-\frac{\sin[p_-(x^--y^-)]}{p_-}\nonumber\\
&=\frac{1}{4}\text{sgn}(x^--y^-)\delta(x^\perp-y^\perp)\ .
\end{align}
The derivation with respect to $x^+$, instead, must be done before the evaluation at $y^+=x^+$. It leads to:
\begin{align}
\partial_+&\left.D(x-y)\right|_{y^+=x^+}\nonumber\\
&=(2\pi)^{-3}\int d^4p\text{sgn}(p_-)p_+\delta(p^2-m^2)\left.e^{-ip(x-y)}\right|_{y^+=x^+}\nonumber\\
&=(2\pi)^{-3}\int d^3\bm p\frac{1}{4p_-^2}(p_\perp^2+m^2)\nonumber\\
&\quad\times e^{-ip_-(x^--y^-)-ip_\perp(x^\perp-y^\perp)}\nonumber\\
&=-\frac{1}{4\partial_-^2}\delta(x^--y^-)(-\partial_\perp^2+m^2)\delta(x^\perp-y^\perp)\nonumber\\
&=-\frac{1}{8}|x^--y^-|(-\partial_\perp^2+m^2)\delta(x^\perp-y^\perp)\ .\label{eq:2.3.22}
\end{align}
Therefore:
\begin{align}
&\left\lbrace\psi(x^+;\bm x);\overline\psi(x^+;\bm y)\right\rbrace=\frac{1}{2}\left\{\gamma^-\delta(x^--y^-)\right.\nonumber\\
&\qquad+\frac{1}{2}\text{sgn}(x^--y^-)(\gamma^\perp\partial_\perp-im)\nonumber\\
&\qquad\left.-\frac{1}{4}\gamma^+|x^--y^-|(-\partial_\perp^2+m^2)\right\}\delta(x^\perp-y^\perp)\ .\label{eq:2.3.23}
\end{align}
This result coincides with the one obtained by Kogut and Soper \cite{KogutSoper}, by Rohrlich and Neville \cite{Rohrlich2, Rohrlich3}, and by the use of Dirac-Bergmann's method via the correspondence principle in Ref. \cite{CPZ}.

The photon quantized field operator in the null-plane gauge, on the other hand, is:
\begin{align}\label{eq:2.4.11}
A^a&(x)=(2\pi)^{-3/2}\nonumber\\
&\times\sum_{\lambda=1,2}\int d\mu(\bm p)\varepsilon_\lambda(\bm p)^a\left(a_\lambda(\bm p)e^{-ipx}+a^\dagger_\lambda(\bm p)e^{ipx}\right)\ .
\end{align}
As we can see, only the physical degrees of freedom --the transversal polarization vectors-- appear in it. For completeness we give the expression of the four polarization vectors, which can be found in a classical analysis \cite{AGPZ}:
\begin{equation*}
\varepsilon_1(\bm p)^a=\left(0;1;0;-\frac{p_1}{p_-}\right)\ ,\  \varepsilon_2(\bm p)^a=\left(0;0;1;-\frac{p_2}{p_-}\right)\ ,
\end{equation*}
\begin{equation}\label{eq:2.4.10.1}
\varepsilon_+(\bm p)^a=\left(1;-\frac{p_1}{p_-};-\frac{p_2}{p_-};\frac{p_\perp^2}{2p_-^2}\right)\ ,\  \varepsilon_-(\bm p)^a=\left(0;0;0;1\right)\ .
\end{equation}
The photon emission and absorption field operators satisfy the following commutation relations:
\begin{equation}\label{eq:2.4.12}
\left[a_\lambda(\bm p);a_\sigma^\dagger(\bm q)\right]=2p_-\delta_{\lambda\sigma}\delta(\bm p-\bm q)\ .
\end{equation}
The commutation distribution for this field operator is:
\begin{equation}\label{eq:2.4.13}
\left[A^a(x);A^b(y)\right]=:iD^{ab}(x-y)\ ,
\end{equation}
with:
\begin{align}\label{eq:2.4.14}
 D^{ab}(x)=&i(2\pi)^{-3}\int d^4pe^{-ipx}\nonumber\\
 &\times\text{sgn}\left(p_-\right)\delta(p^2)\left(g^{ab}-\frac{p^a\eta^b+\eta^ap^b}{p_-}\right)\ .
\end{align}
Here, the vector $\eta$ has components $(\eta^a)=(0;0_\perp;1)$. We can also obtain the restriction of these commutation relations to the null plane $y^+=x^+$. The equal-time commutators between the transversal components of the quantized field operators are:
\begin{align}
&\left[A_\alpha(x^+;\bm x);A_\beta(x^+;\bm y)\right]\nonumber\\
&\qquad=-\frac{i}{4}\delta^\alpha_\beta\text{sgn}(x^--y^-)\delta(x^\perp-y^\perp)\ .\label{eq:2.4.16}
\end{align}
Again, these results agree with the ones in Refs. \cite{Rohrlich2, Rohrlich3, CPZ}.

\section{Null-plane causal perturbation theory}\label{sec:1}

In the causal theory one uses the operation of ``adiabatic switching'' \cite{Bogo}, by means of which the coupling constant of the interaction theory is multiplied by a ``switching function'' $g\in\mathscr{S}(\mathbb{R}^4):\mathbb{R}^4\to\mathbb{R}$, in order to isolate the problem of infra-red divergences, and with it, the problem of the confinement in the real (physical) asymptotic states; it is through the adiabatic limit $g\to 1$ that the real interaction is recovered. This operation allows the usage of the free fields for the construction of the $S(g)$ scattering operator, which is subjected to Bogoliubov-Medvedev-Polivanov's axioms \cite{BogoMedPoli, Bogo, BogoLogunov}: (i) translation invariance, (ii) causality --now referred to the $x^+$ null-plane time--, (iii) unitarity, (iv) Lorentz's invariance and (v) vacuum stability. For the construction of CPT only the axioms (i) and (ii) are necessary, while (iii), (iv) and (v) are physical conditions imposed for the normalization of the scattering operator. The details of the formulation of this theory on light-front dynamics can be found in Ref. \cite{APS22}.

Being a perturbation theory, in CPT the $S(g)$ operator is written as the following formal series:
\begin{equation}\label{eq:3.2.3}
S(g)=1+\sum\limits_{n=1}^{+\infty}\frac{1}{n!}\int dXT_n(X)g(X)\ ,
\end{equation}
with the notations: $T_n(X)\equiv T_n(x_1;\ldots;x_n)$, $g(X)\equiv g(x_1)\ldots g(x_n)$, $dX\equiv d^4x_1\ldots d^4x_n$. Eq. \eqref{eq:3.2.3} is also the definition of the ``transition distributions of order $n$'' or ``$n$-point distributions'' $T_n(x_1;\ldots;x_n)\in\mathscr S'(\mathbb{R}^{4n})$, which are symmetrical in the coordinates $x_1,\ldots,x_n$ as the products of functions $g(x_1)\ldots g(x_n)$ are.

The inverse operator $S(g)^{-1}$ is given as a perturbation series as well:
\begin{equation*}
S(g)^{-1}=1+\sum\limits_{n=1}^{+\infty}\frac{1}{n!}\int dX\widetilde T_n(X)g(X)\ ,
\end{equation*}
\begin{equation}\label{eq:3.2.4}
\widetilde T_n(X)=\sum\limits_{r=1}^n(-1)^r\sum\limits_{\begin{subarray}{l} X_1,\ldots,X_r\neq\emptyset\\ X_1\cup\ldots\cup X_r=X\\ X_j\cap X_k=\emptyset, \forall j\neq k \end{subarray}}T_{n_1}(X_1)\ldots T_{n_r}(X_r)\ .
\end{equation}

The causality axiom implies that the transition distributions are ``chronologically ordered'' --in the sense of the $x^+$ time--:
\begin{align}\label{eq:3.2.17}
&T_n(X)=T_m(X_2)T_{n-m}(X_1)\quad\text{for}\quad X_1<X_2\ ;\nonumber\\
&\left[T_n(X);T_m(Y)\right]=0\quad\text{for}\quad X\sim Y\ .
\end{align}
Because of this, we can define the ``advanced distribution of order $n$'' as the following distribution:
\begin{equation}\label{eq:3.3.2}
A_n(Y;x_n)=\sum\limits_{\begin{subarray}{l} X\cup X'=Y\\ X\cap X'=\emptyset\end{subarray}}\widetilde T_m(X)T_{n-m}(X'\cup\left\{x_n\right\})\quad,
\end{equation}
and the ``retarded distribution of order $n$'' as:
\begin{equation}\label{eq:3.3.4}
R_n(Y;x_n)=\sum\limits_{\begin{subarray}{l} X\cup X'=Y\\ X\cap X'=\emptyset\end{subarray}}T_{n-m}(X'\cup\left\{x_n\right\})\widetilde T_m(X)\quad.
\end{equation}
In these distributions the $n$-point distribution appears once. Separating it from the other terms:
\begin{align}\label{eq:3.3.5}
&A_n(Y;x_n)=T_n(Y\cup\left\{x_n\right\})+A'_n(Y;x_n)\ ,\nonumber\\
&R_n(Y;x_n)=T_n(Y\cup\left\{x_n\right\})+R'_n(Y;x_n)\ ,
\end{align}
with the following definitions of the ``advanced subsidiary distribution'' and of the ``retarded subsidiary distribution'', respectively, which do not contain $T_n$, but only the transition distributions $T_m$ with $m\leq n-1$:
\begin{equation}\label{eq:3.3.6}
A'_n(Y;x_n):=\sum\limits_{\begin{subarray}{l} X\cup X'=Y\\ X\cap X'=\emptyset\\ X\neq\emptyset\end{subarray}}\widetilde T_m(X)T_{n-m}(X'\cup\left\{x_n\right\})\ ,
\end{equation}
\begin{equation}\label{eq:3.3.7}
R'_n(Y;x_n):=\sum\limits_{\begin{subarray}{l} X\cup X'=Y\\ X\cap X'=\emptyset\\ X\neq\emptyset\end{subarray}}T_{n-m}(X'\cup\left\{x_n\right\})\widetilde T_m(X)\ .
\end{equation}
The transition distribution of order $n$ is then equal to --a similar formula holds with the advanced distribution--:
\begin{align}\label{eq:3.3.8}
T_n(Y\cup\left\{x_n\right\})&=R_n(Y;x_n)-R'_n(Y;x_n)\ .
\end{align}
Therefore, the $n$-point distribution can be found by encountering the retarded distribution of order $n$, which can be done by splitting \cite{Division1,Division2,Division3} the ``causal distribution of order $n$'':
\begin{align}\label{eq:3.3.9}
D_n(Y;x_n)&:=R_n(Y;x_n)-A_n(Y;x_n)\nonumber\\
&=R'_n(Y;x_n)-A'_n(Y;x_n)\ .
\end{align}
It must be done as follows: Suppose that the causal distribution of order $n$ was already constructed by means of the inductive procedure; it has, in general, the following form:
\begin{equation}\label{eq:3.5.1}
D_n(x_1;\ldots;x_n)=\sum_kd^k_n(x_1;\ldots;x_n)\normord{C_k(u^A)}\ ,
\end{equation}
with $d^k_n(x_1;\ldots;x_n)$ a numerical distribution and $\normord{C_k(u^A)}$ a Wick's monomial of the different quantized free field operators $u^A$ of the theory. Since these field operators do not restrict the support of the complete distribution, it is sufficient to consider the splitting of the numerical distribution $d^k_n$, whose support, then, is causal by hypothesis. Also the advanced and retarded distributions will maintain the operator fields structure of the causal distribution:
\begin{equation}\label{eq:3.5.2}
A_n(x_1;\ldots;x_n)=\sum_ka^k_n(x_1;\ldots;x_n)\normord{C_k(u^A)}\ ,
\end{equation}
\begin{equation}\label{eq:3.5.3}
R_n(x_1;\ldots;x_n)=\sum_kr^k_n(x_1;\ldots;x_n)\normord{C_k(u^A)}\ ,
\end{equation}
with $a^k_n$ and $r^k_n$ the advanced and retarded parts, respectively, of the numerical distribution $d^k_n$. Using the translational invariance, define the numerical distribution $d\in\mathscr{S}'(\mathbb{R}^{4n-4})$ as:
\begin{align}\label{eq:3.5.6}
&d(x):=d^k_n(x_1-x_n;\ldots;x_{n-1}-x_n;0)\ ,
\end{align}
with $\text{supp}(d)\subseteq\Gamma^+_{n-1}(0)\cup\Gamma^-_{n-1}(0)$, and which will be split as:
\begin{equation}\label{eq:3.5.7}
d=r-a\ ;\ \text{supp}(r)\subseteq\Gamma^+_{n-1}(0)\ ,\ \text{supp}(a)\subseteq\Gamma^-_{n-1}(0)\ .
\end{equation}
Here we are denoting:
\begin{align*}
&\Gamma^+_n(0):=\Big\{(x_1;\cdots;x_n)\in\mathbb{M}^n\ \Big|\ \forall j\in\left\{1,\cdots, n\right\}\ :\nonumber\\
& x_j^+\geq 0\ \wedge\ \left( \exists x_k\in\overline{V^+}(0) (k\neq j) : x_j\in\widetilde V^+(x_k)\right)\Big\}\ ,
\end{align*}
with $V^\pm(x)$ the interior of the future or past, respectively, light-cone with vertex at the point $x$, $\overline{V^\pm}(x)$ its closure and $\widetilde V^\pm(x)$ the union of its closure and the $x^-$-axis. An analogous definition holds for $\Gamma^-_n(0)$. Additionally, in Eq. \eqref{eq:3.5.6} we have written $d(x)$; $x$ means: $(x_1-x_n;\ldots;x_{n-1}-x_n)$. In the following we will use Schwartz's multi-index notation \footnote{It is defined, for example, in Ref. \cite{SchwartzM}: A multi-index $k\in\mathbb{R}^N$ is a sequence of non-negative numbers, $k=(k_1;\ldots;k_N)$, $k_j\geq0$, for which the following notations are established:
\begin{align*}
&|k|\equiv\sum\limits_{j=1}^N k_j\ ,\ x^k\equiv\prod\limits_{j=1}^Nx_j^{k_j}\ ,\  k!\equiv\prod\limits_{j=1}^Nk_j!\ ,\\
&D^kf(x)\equiv\prod\limits_{j=1}^N\partial_{x_j}^{k_j}f(x)\ .
\end{align*}}. We will also use the notation: $x^a\equiv(x_1^a-x_n^a;\ldots;x_{n-1}^a-x_n^a)$.

To perform the splitting, it is crucial to remember that the product of a distribution with a discontinuous function can be ill-defined if the distribution has a singularity precisely on the discontinuity surface of the function \footnote{This was exemplified in Ref. \cite{AP2} with the distribution $\delta$ and the function $\Theta$.}. In our case we then need to control the behaviour of the causal distribution near the splitting region. In instant dynamics, in which the splitting region is the vertex of the light-cone, the concept of Vladimirov-Drozzinov-Zavialov's quasi-asymptotics \cite{Vladimirov} was introduced to cast that behaviour \cite{ScharfBorel}. In null-plane dynamics, the splitting region is the intersection of the null plane $x^+=0$ with the light-cone, which is the entire $x^-$-axis, hence the concept of  quasi-asymptotics by selected variable\footnote{Also developed by Vladimirov, Drozzinov and Zavialov in Ref. \cite{Vladimirov}.} is most adequate for this purpose:

\textbf{Definition:} \textit{Let $d\in\mathscr{S}'(\mathbb{R}^m)$ be a distribution, and let $\rho$ be a continuous positive function. If the (distributional) limit}
\begin{equation}\label{eq:3.5.9}
\lim_{s\to 0^+}\rho(s)s^{3m/4}d\left(sx^+;sx^\perp;x^-\right)=d_-(x)
\end{equation}
\textit{exists in $\mathscr{S}'(\mathbb{R}^m)$ and is non-null, then the distribution $d_-$ is called the \underline{quasi-asymptotics} of $d$ at the $x^-$-axis, with regard to the function $\rho$.}

The function $\rho(s)$ can be shown to be a ``regularly varying at zero function'', also called an ``auto-model function'' \cite{Vladimirov, ScharfFQED}, which means that for every $a>0$: $\displaystyle{\lim_{s\to0^+}\rho(as)/\rho(s)=a^\alpha}$ for some $\alpha\in\mathbb{R}$, called the ``order of auto-modelity'' of the function $\rho$. This number serves as a characterizing parameter of the distribution, which is called its ``singular order at the $x^-$-axis'' and is denoted by $\omega_-$.

In momentum space the following splitting \textit{formulae} are found: For negative singular order, $\omega_-<0$:
\begin{align}
\hat r(p)&=\frac{i}{2\pi}\int\limits_{-\infty}^{+\infty} \frac{\hat d(p_+-k;\bm p)}{k+i0^+}dk\ .\label{eq:3.6.9}
\end{align}
For non-negative singular order, $\omega_-\geq0$, the ``retarded distribution with normalization line $\left(q_+;q_\perp;p_-\right)$'' is:
\begin{align}\label{eq:3.6.47}
\hat r_q&(p)=\frac{i}{2\pi}\int\limits_{-\infty}^{+\infty}\frac{dk}{k+i0^+}\Bigg\{\hat d(p_+-k;\bm p)\nonumber\\
&-\sum\limits_{|c|=0}^{\lfloor\omega_-\rfloor}\frac{1}{c!}(p_{+,\alpha}-q_{+,\alpha})^cD^c_{+,\alpha}\hat d(q_+-k;q_\perp;p_-)\Bigg\}\ .
\end{align}
A particular case of normalization line is $(0;0_\perp;p_-)$; the solution normalized at it is called the ``central solution''.

Finally, if $r_1$ and $r_2$ are two solutions of the splitting problem, then they could be different only by ``normalization terms'' which are distributions with support on the $x^-$-axis. In momentum space:
\begin{equation}\label{eq:3.7.5}
\hat r_1(p)-\hat r_2(p)=\sum\limits_{|b|=0}^{M}\widehat C_b\left(p_-\right)p_{+,\perp}^b\ ,
\end{equation}
with $\widehat C_b\left(p_-\right)$ some distributions of the variable $p_-$. The singular order of each one of these terms is $|b|$, independently of which the distribution $\widehat C_b(p_-)$ is, because the variable $p_-$ is not scaled in the singular order calculus in light-front dynamics; this leads to a richer variety of possible normalization terms when compared to instant dynamics; particularly, instantaneous normalization terms are now allowed. The procedure of determining these unknown distributions by the imposition of physical requirements is called the ``normalization process''.

\section{Causal distribution of the second order QED}\label{sec:2}

For QED the first order term of the $S(g)$ operator is given by the one-point distribution:
\begin{equation}\label{eq:8.2.1}
T_1(x)=i\normord{j^a(x)}A_a(x)\equiv ie\normord{\overline{\psi}(x)\gamma^a\psi(x)}A_a(x)\ .
\end{equation}
The construction of the second order causal distribution starts with the definition of the subsidiary ones:
\begin{equation}\label{eq:8.2.2}
A'_2(x_1;x_2)=\widetilde{T}_1(x_1)T_1(x_2)=-T_1(x_1)T_1(x_2)\ ,
\end{equation}
\begin{equation}\label{eq:8.2.3}
R'_1(x_1;x_2)=T_1(x_2)\widetilde{T}_1(x_1)=-T_1(x_2)T_1(x_1)\ ,
\end{equation}
with which the causal distribution $D_2=R'_2-A'_2$ is equal to:
\begin{equation}\label{eq:8.2.4}
D_2(x_1;x_2)=\left[T_1(x_1);T_1(x_2)\right]\ .
\end{equation}
The explicit expression of this distribution is obtained by replacing Eq. \eqref{eq:8.2.1} into Eq. \eqref{eq:8.2.4}, and by using Wick's theorem with the contractions:
\begin{align}
\wick{\c1\psi_a(x)\c1{\overline{\psi}}_b(y)}&=\frac{1}{i}S_{ab+}(x-y)\ ,\label{eq:8.2.5}\\
\wick{\c1{\overline{\psi}}_a(x)\c1\psi_b(y)}&=\frac{1}{i}S_{ba-}(y-x)\ ,\label{eq:8.2.6}\\
\wick{\c A_a(x)\c A_b(y)}&=iD_{ab+}(x-y)\ .\label{eq:8.2.7}
\end{align}

We will need the subsidiary retarded distribution --see Eq. \eqref{eq:3.3.8}--, given by:
\begin{align}
R'_2&(x_1;x_2)=e^2\gamma^{a}_{ab}\gamma^{b}_{cd}\left[\normord{\overline{\psi}_a(x_1)\psi_b(x_1)\overline{\psi}_c(x_2)\psi_d(x_2)}\right.\nonumber\\
&+iS_{bc-}(x_1-x_2)\normord{\overline{\psi}_a(x_1)\psi_d(x_2)}\nonumber\\
&+iS_{da+}(x_2-x_1)\normord{\psi_b(x_1)\overline{\psi}_c(x_2)}\nonumber\\
&\left.-S_{bc-}(x_1-x_2)S_{da+}(x_2-x_1)\right]\nonumber\\
&\times\left[\normord{A_a(x_1)A_b(x_2)}+iD_{ab+}(x_2-x_1)\right]\ ,\label{eq:8.2.9}
\end{align}
and the causal distribution, whose final form is:
\begin{align}\label{eq:8.2.10}
D_2=D_2^{(M)}+D_2^{(C)}+D_2^{(VP)}+D_2^{(SE)}+D_2^{(VG)}\ ,
\end{align}
with --we use the relative coordinate $y\equiv x_1-x_2$--:
\begin{align}
&D_2^{(M)}(x_1;x_2)=-ie^2D_{ab}(y)\nonumber\\
&\times\normord{\overline\psi(x_1)\gamma^{a}\psi(x_1)\overline\psi(x_2)\gamma^{b}\psi(x_2)}\ ,\label{eq:8.2.11}
\end{align}
\begin{align}
&D_2^{(C)}(x_1;x_2)=ie^2\normord{A_a(x_1)A_b(x_2)}\nonumber\\
&\times\big[\normord{\overline\psi(x_1)\gamma^{a}S(y)\gamma^{b}\psi(x_2)}\nonumber\\
&-\normord{\overline\psi(x_2)\gamma^{b}S(-y)\gamma^{a}\psi(x_1)}\big]\ ,\label{eq:8.2.12}
\end{align}
\begin{align}
&D_2^{(VP)}(x_1;x_2)=-e^2\normord{A_{a}(x_1)A_{b}(x_2)}\nonumber\\
&\times\text{Tr}\left[\gamma^{a}S_-(y)\gamma^{b}S_+(-y)-\gamma^{a}S_+(y)\gamma^{b}S_-(-y)\right]\ ,\label{eq:8.2.13}
\end{align}
\begin{align}
&D_2^{(SE)}(x_1;x_2)=-e^2\normleft{\overline\psi(x_1)}\gamma^{a}\big[S_-(y)D_{ab+}(-y)\nonumber\\
&+S_+(y)D_{ab+}(y)\big]\gamma^{b}\normright{\psi(x_2)}+e^2\normleft{\overline\psi(x_2)}\gamma^{a}\nonumber\\
&\times\big[S_+(-y)D_{ab+}(-y)+S_-(-y)D_{ab+}(y)\big]\gamma^{b}\normright{\psi(x_1)}\ ,\label{eq:8.2.14}
\end{align}
\begin{align}
&D_2^{(VG)}(x_1;x_2)=-ie^2D_{ab+}(-y)\nonumber\\
&\times\text{Tr}\left[\gamma^{a}S_-(y)\gamma^{b}S_+(-y)-\gamma^{a}S_-(-y)\gamma^{b}S_+(y)\right]\ .\label{eq:8.2.15}
\end{align}
In this form, Eqs. \eqref{eq:8.2.11}-\eqref{eq:8.2.15} allow to directly identify the terms which will contribute to each process: the non-contracted quantized field operators determine the initial and final states which will give a non-null contribution to the amplitude $\left\vert\left<\Psi;S(g)\Phi\right>\right\vert^2$. Hence, the distribution $D_2^{(M)}$ describes the scattering of two fermions, $D^{(C)}_2$, the scattering of a fermion by a photon; the distributions $D_2^{(VP)}$ and $D_2^{(SE)}$ represent the vacuum polarization and fermion's self-energy, respectively; finally, the distribution $D_2^{(VG)}$ do not describe any physical process.

\section{Scattering processes}\label{sec:3}

In this section we will show in a very direct manner that the equivalence with instant dynamics for the scattering processes at second order can be obtained by a suitable choice of the normalization terms.

\subsection{M{\o}ller's scattering}

The scattering of two leptons, called M{\o}ller's scattering, is described by the causal distribution in Eq. \eqref{eq:8.2.11}. The numerical distribution contained in it is the commutation distribution of the radiation field, whose expression in momentum space and in the null-plane gauge is:
\begin{equation}\label{eq:3.6.20}
\widehat D_{ab}(p)=\frac{i}{2\pi}\text{sgn}\left(p_-\right)\delta(p^2)\left(g_{ab}-\frac{p_a\eta_b+\eta_ap_b}{p_-}\right)\ .
\end{equation}
Some of the components of this causal distribution have singular order $\omega_-=-2$, while others have $\omega_-=-1$. In any case, the singular order is negative, so the retarded part is found by application of Eq. \eqref{eq:3.6.9}. In order to do that, it will be convenient to define the following distributions:
\begin{equation}\label{eq:3.6.21}
\hat d_1(p):=\frac{i}{2\pi}\text{sgn}\left(p_-\right)\delta(p^2)\ ,\ \hat d_{2a}(p):=\hat d_1(p)\frac{p_a}{p_-}\ ;
\end{equation}
as a function of which the commutation distribution is:
\begin{equation}\label{eq:3.6.22}
\widehat D_{ab}(p)=g_{ab}\hat d_1(p)-\left[\hat d_{2a}(p)\eta_{b}+\eta_{a}\hat d_{2b}(p)\right]\ .
\end{equation}
To find the retarded part of $\widehat{D}_{ab}$ is then equivalent to find the retarded parts of $\hat d_1$ and $\hat d_{2a}$. We find, by using the variable $s=-2kp_-$:
\begin{align}
\hat r_1(p)&=-(2\pi)^{-2}\int\limits_{-\infty}^{+\infty}\frac{\text{sgn}(p_-)\delta(p^2-2kp_-)}{k+i0^+}dk\nonumber\\
&=(2\pi)^{-2}\int\limits_{-\infty}^{+\infty}\frac{\delta(s+p^2)}{s-ip_-0^+}ds\nonumber\\
&=-(2\pi)^{-2}\frac{1}{p^2+ip_-0^+}\ ;\label{eq:3.6.23}
\end{align}
since the variables $p_{\alpha,-}$ do not change in the splitting formula of Eq. \eqref{eq:3.6.9}:
\begin{align}
\hat r_{2\alpha,-}(p)&=\frac{p_{\alpha,-}}{p_-}\hat r_1(p)=-(2\pi)^{-2}\frac{1}{p^2+ip_-0^+}\frac{p_{\alpha,-}}{p_-}\ ;\label{eq:3.6.24}
\end{align}
and finally:
\begin{align}
\hat r_{2+}(p)&=-(2\pi)^{-2}\int\limits_{-\infty}^{+\infty}\frac{\text{sgn}(p_-)\delta(p^2-2kp_-)(p_+-k)}{p_-(k+i0^+)}dk\nonumber\\
&=(2\pi)^{-2}\int\limits_{-\infty}^{+\infty}\frac{\delta(s+p^2)\left(p_++\dfrac{s}{2p_-}\right)}{p_-(s-ip_-0^+)}ds\nonumber\\
&=-(2\pi)^{-2}\left(\frac{1}{p^2+ip_-0^+}\frac{p_+}{p_-}-\frac{1}{2p_-}\right)\ .\label{eq:3.6.25}
\end{align}
Eqs. \eqref{eq:3.6.23}-\eqref{eq:3.6.25} imply that the retarded part of the commutation distribution of the massless vector field is [see Eq. \eqref{eq:3.6.22}]:
\begin{align}\label{eq:3.6.26}
&\widehat D^{\text{ret}}_{ab}(p)=-\frac{(2\pi)^{-2}}{p^2+ip_-0^+}\nonumber\\
&\times\left\lbrace g_{ab}-\frac{p_a\eta_{b}+\eta_{a}p_b}{p_-}+\frac{p^2}{2p_-^2}\left[\delta_{a+}\eta_{b}+\eta_{a}\delta_{b+}\right]\right\rbrace\ .
\end{align}
But, since $\delta_{a+}$ are precisely the components of $\eta_{a}$, the above equation simplifies to:
\begin{equation}\label{eq:3.6.27}
\widehat D^{\text{ret}}_{ab}(p)=\frac{-(2\pi)^{-2}}{p^2+ip_-0^+}\left\lbrace g_{ab}-\frac{p_a\eta_{b}+\eta_{a}p_b}{p_-}+\frac{p^2}{p_-^2}\eta_{a}\eta_{b}\right\rbrace\ .
\end{equation}
Subtracting the subsidiary retarded distribution, which corresponds to the negative frequency part of the commutation distribution,
\begin{equation}\label{eq:3.6.28}
\widehat D_{ab-}(p)=-\frac{i}{2\pi}\Theta(-p_-)\delta(p^2)\left\lbrace g_{ab}-\frac{p_a\eta_{b}+\eta_{a}p_b}{p_-}\right\rbrace\ ,
\end{equation}
we obtain Feynman's propagator of this quantized field:
\begin{align}\label{eq:3.6.29}
&\widehat D^{F}_{ab}(p):=\widehat D_{ab}^{\text{ret}}(p)-\widehat D_{ab-}(p)\nonumber\\
&=-\frac{(2\pi)^{-2}}{p^2+i0^+}\left\lbrace g_{ab}-\frac{p_a\eta_{b}+\eta_{a}p_b}{p_-}+\frac{p^2}{p_-^2}\eta_{a}\eta_{b}\right\rbrace\ ,
\end{align}
which is the one that enters into the transition distribution for M{\o}ller's scattering:
\begin{align}\label{eq:8.3.1}
T_2^{(M)}&(x_1;x_2)=-ie^2D^F_{ab}(y)\nonumber\\
&\times\normord{\overline\psi(x_1)\gamma^{a}\psi(x_1)\overline\psi(x_2)\gamma^{b}\psi(x_2)}\ ,
\end{align}
As we see in Eq. \eqref{eq:3.6.27}, we have obtained an instantaneous term in the splitting process of the causal distribution. This has led to the so-called ``doubly transverse gauge propagator'', shown in Eq. \eqref{eq:3.6.29}, which means that $\widehat D^F_{ab}(p)$ is transverse both to $p^a$ and $\eta^a$ \cite{TenEyck1,TenEyck2,Prem2}; here we have obtained it in a very natural way. Now, the singular order of this distribution is $\omega_-\left[D^F_{ab}\right]=0$, so it is allowed a normalization term of the form $\widehat C\left(p_{-}\right)$. Choosing:
\begin{equation}\label{eq:8.3.3}
\widehat C\left(p_{-}\right)=-ie^2(2\pi)^{-2}\frac{\eta_{a}\eta_{b}}{p_{-}^2}\ ,
\end{equation}
the instantaneous term which arose in the splitting procedure cancels out and we are left with:
\begin{align}\label{eq:8.3.4}
\hat d_{ab}(p)&\equiv \widehat D^F_{ab}(p)+\widehat C\left(p_{-}\right)\nonumber\\
&=-\frac{(2\pi)^{-2}}{p^2+i0^+}\left\{g_{ab}-\frac{p_{a}\eta_{b}+\eta_{a}p_{b}}{p_{-}}\right\}\ .
\end{align}

Consider now the following initial and final states, respectively, with definite \textit{momenta}:
\begin{equation}\label{eq:8.3.5}
\frac{b_{s_1}^\dagger(\bm p_1)}{\sqrt{2p_{1-}}}\frac{b_{r_1}^\dagger(\bm q_1)}{\sqrt{2q_{1-}}}\Omega\ ,\ \frac{b_{s_2}^\dagger(\bm p_2)}{\sqrt{2p_{2-}}} \frac{b_{r_2}^\dagger(\bm q_2)}{\sqrt{2q_{2-}}}\Omega\ .
\end{equation}
For these states, the $S$ operator to second order in the adiabatic limit $g\to1$ is:
\begin{align}
&S_{12}^{(M)}=\frac{1}{\sqrt{2p_{1-}2q_{1-}2p_{2-}2q_{2-}}}\nonumber\\
&\quad\times\left(\Omega;b_{r_2}(\bm q_2)b_{s_2}(\bm p_2)S_2^{(M)}b_{s_1}(\bm p_1)^\dagger b_{r_1}(\bm q_1)^\dagger\Omega\right)\nonumber\\
&=-\frac{1}{\sqrt{2p_{1-}2q_{1-}2p_{2-}2q_{2-}}}\frac{ie^2}{2(2\pi)^2}\int d^4kd^4x_1d^4x_2e^{-iky}\nonumber\\
&\ \times\hat d_{ab}(k)\big(\Omega;b_{r_2}(\bm q_2)b_{s_2}(\bm p_2)\nonumber\\
&\ \times\normord{\overline\psi(x_1)\gamma^{a}\psi(x_1)\overline\psi(x_2)\gamma^{b}\psi(x_2)} b_{s_1}(\bm p_1)^\dagger b_{r_1}(\bm q_1)^\dagger\Omega\big)\ .\label{eq:8.3.6}
\end{align}
Inside the parentheses, the non-null contributions are found by using Wick's theorem. There are four contributions, which we obtain by using the contractions:
\begin{align}\label{eq:8.3.7}
&\wick{\c b_s(\bm p)\c{\overline\psi}(x)}=(2\pi)^{-3/2}\Theta\left(p_{-}\right)\sqrt{2p_-}\overline u_s(\bm p)e^{ipx}\ ,\nonumber\\
&\wick{\c\psi(x)\c b_s(\bm p)^\dagger}=(2\pi)^{-3/2}\Theta\left(p_{-}\right)\sqrt{2p_-}u_s(\bm p)e^{-ipx}\ .
\end{align}
Then:
\begin{align}
&S^{(M)}_{12}=\frac{ie^2}{2(2\pi)^8}\int d^4kd^4x_1d^4x_2 e^{-iky}\hat d_{ab}(k)\Theta\left(p_{1-}\right)\nonumber\\
&\times\Theta\left(q_{1-}\right)\Theta\left(p_{2-}\right)\Theta\left(q_{2-}\right)\Big\{\overline u_{s_2}(\bm p_2)\gamma^{a}u_{r_1}(\bm q_1)\nonumber\\
&\times\overline u_{r_2}(\bm q_2)\gamma^{b}u_{s_1}(\bm p_1)e^{i(p_2-q_1)x_1+i(q_2-p_1)x_2}-\overline u_{s_2}(\bm p_2)\gamma^{a}\nonumber\\
&\times u_{s_1}(\bm p_1)\overline u_{r_2}(\bm q_2)\gamma^{b}u_{r_1}(\bm q_1)e^{i(p_2-p_1)x_1+i(q_2-q_1)x_2}\nonumber\\
&-\overline u_{r_2}(\bm q_2)\gamma^{a}u_{r_1}(\bm q_1)\overline u_{s_2}(\bm p_2)\gamma^{b}u_{s_1}(\bm p_1)\nonumber\\
&\times e^{i(q_2-q_1)x_1+i(p_2-p_1)x_2}+\overline u_{r_2}(\bm q_2)\gamma^{a}u_{s_1}(\bm p_1)\nonumber\\
&\overline u_{s_2}(\bm p_2)\gamma^{b}u_{r_1}(\bm q_1)e^{i(q_2-p_1)x_1+i(p_2-q_1)x_2}\Big\}\ .\label{eq:8.3.8}
\end{align}
In all the terms the following integral appears:
\begin{equation}\label{eq:8.3.9}
\int d^4x_1d^4x_2e^{-iky}e^{iPx_1+iQx_2}=(2\pi)^8\delta(k-P)\delta(P+Q)\ ,
\end{equation}
so that, integrating in the variable $k$ and using the symmetries of the $\hat d_{ab}$ distribution [see Eq. \eqref{eq:8.3.4}],
\begin{equation}\label{eq:8.3.10}
\hat d_{ab}(k)=\hat d_{ab}(-k)\quad\text{and}\quad \hat d_{ab}(k)=\hat d_{ba}(k)\ ,
\end{equation}
we finally find:
\begin{align}
&S^{(M)}_{12}=ie^2\delta(p_2+q_2-p_1-q_1)\nonumber\\
&\times\Theta\left(p_{1-}\right)\Theta\left(q_{1-}\right)\Theta\left(p_{2-}\right)\Theta\left(q_{2-}\right)\nonumber\\
&\times\left\{\overline u_{s_2}(\bm p_2)\gamma^{a}u_{r_1}(\bm q_1)\overline u_{r_2}(\bm q_2)\gamma^{b}u_{s_1}(\bm p_1)\hat d_{ab}(p_2-q_1)\right.\nonumber\\
&\left.-\overline u_{s_2}(\bm p_2)\gamma^{a}u_{s_1}(\bm p_1)\overline u_{r_2}(\bm q_2)\gamma^{b}u_{r_1}(\bm q_1)\hat d_{ab}(p_2-p_1)\right\}\ .\label{eq:8.3.11}
\end{align}

The wave-functions $u(p)$ and $\overline u(p)$ satisfy Dirac's equation in momentum space:
\begin{equation}\label{eq:8.3.12}
\slashed pu(p)=mu(p)\quad,\quad\overline u(p)\slashed p=m\overline u(p)\ .
\end{equation}
Therefore:
\begin{align}\label{eq:8.3.13}
&\overline u_{s_2}(\bm p_2)\gamma^{a}u_{r_1}(\bm q_1)\left(p_{2a}-q_{1a}\right)=\nonumber\\
&\quad\overline u_{s_2}(\bm p_2)\left(\slashed p_2-\slashed q_1\right)u_{r_1}(\bm q_1)=0\ ,
\end{align}
and the non-covariant terms in the $\hat d_{ab}$ distribution do not contribute to $S_{12}^{(M)}$ [see Eqs. \eqref{eq:8.3.4} and \eqref{eq:8.3.11}]. We conclude that all the non-local terms cancel out, and the result is the same as if we would consider the covariant part of the radiation field commutation distribution only:
\begin{equation}\label{eq:8.3.14}
-\frac{(2\pi)^{-2}}{k^2+i0^+}g_{ab}\ ,
\end{equation}
establishing the equivalence with instant dynamics.

\subsection{Compton's scattering}

Now we turn to the study of Compton's scattering, this is to say, the scattering of a fermion by a photon, whose causal distribution at second order is the one in Eq. \eqref{eq:8.2.12}. Defining the numerical distribution:
\begin{equation}\label{eq:8.4.1}
d^{ab}(y)=ie^2\gamma^{a}S(y)\gamma^{b}\ ,
\end{equation}
we will have:
\begin{align}\label{eq:8.4.2}
&D_2^{(C)}(x_1;x_2)=\normord{A_{a}(x_1)A_{b}(x_2)}\nonumber\\
&\times\left(\normord{\overline\psi(x_1)d^{ab}(y)\psi(x_2)}-\normord{\overline\psi(x_2)d^{ba}(-y)\psi(x_1)}\right)\ .
\end{align}
The distribution $d^{ab}(y)$ has singular order $\omega_-=-1$, and its retarded part is:
\begin{equation}\label{eq:8.4.3}
\hat r^{ab}(y)=ie^2\gamma^{a}S^{\text{ret}}(y)\gamma^{b}\ .
\end{equation}
Hence we need to obtain the retarded part of the anti-commutation distribution of the fermion field. In momentum space it is:
\begin{equation}\label{eq:3.6.15}
\widehat S(p)=\frac{i}{2\pi}(\slashed p+m)\text{sgn}\left(p_-\right)\delta(p^2-m^2)\ .
\end{equation}
As it was said, its singular order at the $x^-$-axis is $\omega_-=-1<0$, so its retarded part is given by Eq. \eqref{eq:3.6.9}:
\begin{align}
&\widehat S^{\text{ret}}(p)=-(2\pi)^{-2}\int\frac{dk}{k+i0^+}\sgn(p_-)\left[(p_+-k)\gamma^+\right.\nonumber\\
&\left.+p_\perp\gamma^\perp+p_-\gamma^-+m\right]\delta(2p_+p_--2kp_--\omega_p^2)\ .\label{eq:3.6.16}
\end{align}
Using the variable $s=-2kp_-$, the above integral is equal to:
\begin{align}
&\widehat S^{\text{ret}}(p)=(2\pi)^{-2}\int\frac{ds}{s-iq0^+}\delta(s+2p_+p_--\omega_p^2)\nonumber\\
&\quad\times\left(p_+\gamma^++p_\perp\gamma^\perp+p_-\gamma^-+m+\dfrac{s}{2p_-}\gamma^+\right)\nonumber\\
&=-(2\pi)^{-2}\frac{\slashed p+m-\dfrac{2p_+p_--\omega_p^2}{2p_-}\gamma^+}{2p_+p_--\omega_p^2+ip_-0^+}\nonumber\\
&=-(2\pi)^{-2}\left(\frac{\slashed p+m}{p^2-m^2+ip_-0^+}-\frac{\gamma^+}{2p_-}\right)\ .\label{eq:3.6.17}
\end{align}
Subtracting the corresponding $\hat r'^{ab}(y)$ subsidiary distribution, which corresponds to the negative frequency part of the anti-commutation distribution,
\begin{equation}\label{eq:3.6.18}
\widehat S_-(p)=-\frac{i}{2\pi}\Theta\left(-p_-\right)(\slashed p+m)\delta(p^2-m^2)\ ,
\end{equation}
we obtain for the numerical part of the transition distribution:
\begin{equation}\label{eq:8.4.4}
t^{ab}(y)=ie^2\gamma^{a}S^{F}(y)\gamma^{b}\ ,
\end{equation}
with Feynman's propagator being:
\begin{align}\label{eq:3.6.19}
\widehat S^F(p)&:=\widehat S_-(p)-\widehat S^{\text{ret}}(p)\nonumber\\
&=(2\pi)^{-2}\left(\frac{\slashed p+m}{p^2-m^2+i0^+}-\frac{\gamma^+}{2p_-}\right)\ .
\end{align}
As we see, in the splitting process of the causal distribution an instantaneous term arise. Writing the normalization term that is allowed for $\omega_-=0$, which is the singular order of the distribution $t^{ab}$:
\begin{equation}\label{eq:8.4.5}
\hat t^{ab}(p)=\frac{ie^2}{(2\pi)^2}\gamma^{a}\left(\frac{\slashed p+m}{p^2-m^2+i0^+}-\frac{\gamma^{+}}{2p_{-}}\right)\gamma^{b}+\widehat C\left(p_{-}\right)\ .
\end{equation}
Choosing
\begin{equation}\label{eq:8.4.6}
\widehat C\left(p_{-}\right)=\frac{ie^2}{(2\pi)^2}\frac{\gamma^{a}\gamma^{+}\gamma^{b}}{2p_{-}}\ ,
\end{equation}
the instantaneous non-covariant term in the fermion Feynman's propagator is canceled out, and we arrive at the final result:
\begin{equation}\label{eq:8.4.7}
\hat t^{ab}(p)=\frac{ie^2}{(2\pi)^2}\gamma^{a}\frac{\slashed p+m}{p^2-m^2+i0^+}\gamma^{b}\ ,
\end{equation}
showing, also for this scattering process, the equivalence with instant dynamics. 

\subsection{Interaction Lagrangian density}\label{sec:Lag}

In our study of the scattering processes we have seen that Lorentz's covariance requires the introduction of very specific normalization terms. In the case of M{\o}ller's scattering, the contribution of the normalization term in Eq. \eqref{eq:8.3.3} to the second order $S(g)$ operator in the adiabatic limit $g\to 1$ is:
\begin{align}
&+\frac{i}{2}\int d^4x_1d^4x_2\normord{j^a(x_1)\delta(y)\frac{\eta_a\eta_b}{\partial_-^2}j^b(x_2)}\nonumber\\
&=\int d^4x_1\normord{j^+(x_1)\frac{i}{2\partial_-^2}j^+(x_1)}\ .\label{eq:8.3.4-1}
\end{align}
This is precisely the instantaneous term which in the usual approach appears in the Lagrangian density by solving the constraint equation for the radiation field in the null-plane gauge in the interacting theory \cite{TenEyck1,TenEyck2}.

Another normalization term was required to obtain a covariant transition distribution for Compton's scattering --Eq. \eqref{eq:8.4.6}--; its contribution  to the scattering operator in the adiabatic limit $g\to1$ is --taking into account the two terms in Eq. \eqref{eq:8.4.2}--:
\begin{align}\label{eq:8.4.8}
&+\frac{1}{2}\int d^4x_1d^4x_2 e^2\left\{\normleft{\left(\overline\psi(x_1)\gamma^aA_a(x_1)\right)}\right.\nonumber\\
&\quad\times\delta(y)\frac{\gamma^+}{2\partial_-}\normright{\left(\gamma^bA_b(x_2)\psi(x_2)\right)}\nonumber\\
&\quad\left.+\normord{\left(\overline\psi(x_2)\gamma^bA_b(x_2)\right)\delta(y)\frac{\gamma^+}{2\partial_-}\left(\gamma^aA_a(x_1)\psi(x_1)\right)}\right\}\nonumber\\
&=\int d^4x_1e^2\normord{\overline\psi(x_1)\gamma^aA_a(x_1)\frac{\gamma^+}{2\partial_-}\gamma^bA_b(x_1)\psi(x_1)} .
\end{align}
The term so obtained is the one which corresponds, in the Lagrangian approach, to the instantaneous interaction term which arises when solving the constraint equation for the fermion field in the interacting theory \cite{TenEyck1,TenEyck2}.

Joining Eqs. \eqref{eq:8.2.1}, \eqref{eq:8.3.4-1} and \eqref{eq:8.4.8} we can identify the ``interaction Lagrangian density'', defined as $-i$ times the one-point transition distribution plus $-i$ times the contribution of the normalization terms of the next-order transition distributions to the scattering operator in the adiabatic limit:
\begin{align}\label{eq:8.4.9}
&\mathscr L=\normord{j^a(x)A_a(x)}-\normord{\frac{1}{2}\left(\frac{1}{\partial_-}j^+(x)\right)^2}\nonumber\\
&+\frac{e^2}{2}\normord{\left(\overline\psi(x)\gamma^aA_a(x)\right)\frac{\gamma^+}{i\partial_-}\left(\gamma^bA_b(x)\psi(x)\right)}\ .
\end{align}
This Lagrangian density was first obtained by Kogut and Soper \cite{KogutSoper}. We remind the reader that the Lagrangian density of Eq. \eqref{eq:8.4.9} is of first order in $e$ when wrote as a function of interacting fields; its second order structure arises when the constraint equations are solved and reintroduced in it. Therefore, that these terms appear in null-plane CPT at the second order is in accordance with the philosophy of the causal approach, which works with free fields only.

Now, it is a debate question if the instantaneous terms in this Lagrangian density cancel exactly the terms coming from the instantaneous terms in the field propagators. Null-plane CPT answer this question in a direct way: Since the normalization terms cancel the instantaneous terms of the propagators at second order, it will cancel them at all orders in a perturbation series based on $\mathscr L$, because the next-order causal distributions are constructed with the normalized transition distributions. Here we see the advantage of working in an inductive framework. 

\section{Vacuum polarization}\label{sec:vacpol}

We consider in this subsection the radiative correction known as ``vacuum polarization'', which will be precisely defined later on, and which comes from the study of the causal distribution in Eq. \eqref{eq:8.2.13}; we write it as:
\begin{align}
D_2^{(VP)}(x_1;x_2)=\left(P^{ab}(y)-P^{ba}(-y)\right)\normord{A_{a}(x_1)A_{b}(x_2)} ,\label{eq:8.5.1}
\end{align}
with:
\begin{equation}\label{eq:8.5.2}
P^{ab}(y)=e^2\text{Tr}\left[\gamma^{a}S_+(y)\gamma^{b}S_-(-y)\right]\ .
\end{equation}
Fourier's transform of the $P^{ab}$ distribution is:
\begin{align}
&\widehat P^{ab}(k)=e^2(2\pi)^{-2}\int d^4p\text{Tr}\left[\gamma^{a}S_+(p)\gamma^bS_-(p-k)\right]\nonumber\\
&=e^2(2\pi)^{-2}\int d^4p\text{Tr}\left[\gamma^{a}(\slashed p+m)\gamma^{b}(\slashed p-\slashed k+m)\right]\nonumber\\
&\quad\times\widehat D_+(p)\widehat D_-(p-k)\ .\label{eq:8.5.3}
\end{align}
The trace which appears here is calculated by the usual techniques:
\begin{align}
&\text{Tr}\left[\gamma^a(\slashed p+m)\gamma^b(\slashed p-\slashed k+m)\right]\nonumber\\
&=\text{Tr}\left[\gamma^a\slashed p\gamma^b(\slashed p-\slashed k)\right]+m^2\text{Tr}\left[\gamma^a\gamma^b\right]\nonumber\\
&=4\left[p^a\left(p^b-k^b\right)+p^{b}\left(p^{a}-k^{a}\right)-g^{ab}\left(p(p-k)-m^2\right)\right] .\label{eq:8.5.4}
\end{align}
Also, since:
\begin{equation}\label{eq:8.5.5}
\widehat D_\pm(p)=\pm\frac{i}{2\pi}\Theta\left(\pm p_{-}\right)\delta(p^2-m^2)\ ,
\end{equation}
in the integrand of Eq. \eqref{eq:8.5.3} the following Dirac's delta distributions will appear: $\delta(p^2-m^2)$ and $\delta((p-k)^2-m^2)$; they imply that $p^2=m^2$ and $k^2=2pk$. Hence:
\begin{align}
\widehat P^{ab}&(k)=\frac{4e^2}{(2\pi)^4}\int d^4p\left[2p^{a}p^{b}-p^{a}k^{b}-k^{a}p^{b}+g^{ab}pk\right]\nonumber\\
&\times\Theta\left(p_-\right)\delta(p^2-m^2)\Theta\left(k_--p_-\right)\delta(k^2-2pk)\ .\label{eq:8.5.6}
\end{align}
We can see from Eq. \eqref{eq:8.5.6} that the distribution $\widehat P^{ab}(k)$ is symmetric in its indices. And, in addition, it is orthogonal to the momentum $k$, since the multiplication by this vector gives:
\begin{equation}\label{eq:8.5.7}
k_{a}\widehat P^{ab}(k)\sim (2pk-k^2)p^{b}\ ,
\end{equation}
which is null by means of the support of $\delta(k^2-2pk)$. Therefore, $\widehat P^{ab}(k)$ must be proportional to the projector $k^{a}k^{b}-k^2g^{ab}$:
\begin{equation}\label{eq:8.5.8}
\widehat P^{ab}(k)=\left(k^{a}k^{b}-k^2g^{ab}\right)B(k^2)\ .
\end{equation}
Taking the trace of this equation and also in Eq. \eqref{eq:8.5.6}, we obtain the following formula for $B(k^2)$:
\begin{equation}\label{eq:8.5.9}
B(k^2)=-\frac{1}{3k^2}\widehat P\indices{^a_{a}}(k)=-\frac{4e^2}{3(2\pi)^4}\left(1+\frac{2m^2}{k^2}\right)I(k)\ ,
\end{equation}
with $I(k)$ the following integral
\begin{equation}\label{eq:8.5.9-1}
I(k)=\int d^4p\Theta\left(p_-\right)\Theta\left(k_--p_-\right)\delta(p^2-m^2)\delta(k^2-2pk)\ .
\end{equation}
This integral is the same which appears in the calculus of boson's self-energy in Yukawa's model --see Ref. \cite{AP2}--. It is equal to:
\begin{equation}\label{eq:8.5.9-2}
I(k)=\frac{\pi}{2}\Theta\left(k_-\right)\Theta(k^2-4m^2)\sqrt{1-\frac{4m^2}{k^2}}\ .
\end{equation}
Then:
\begin{align}\label{eq:8.5.10}
&\widehat P^{ab}(k)=\frac{e^2}{3(2\pi)^3}\left(g^{ab}-\frac{k^{a}k^{b}}{k^2}\right)(k^2+2m^2)\Theta\left(k_{-}\right)\nonumber\\
&\times\Theta\left(k^2-4m^2\right)\sqrt{1-\frac{4m^2}{k^2}}\ .
\end{align}
The numerical distribution associated with vacuum-polarization is therefore [see Eq. \eqref{eq:8.5.1}]:
\begin{align}\label{eq:8.5.11}
&\hat d^{ab}(k)=\frac{e^2}{3(2\pi)^3}\left(g^{ab}-\frac{k^{a}k^{b}}{k^2}\right)(k^2+2m^2)\text{sgn}\left(k_{-}\right)\nonumber\\
&\quad\times\Theta\left(k^2-4m^2\right)\sqrt{1-\frac{4m^2}{k^2}}\ .
\end{align}

In order to obtain its retarded part we will factorize a second order polynomial:
\begin{equation}\label{eq:8.5.12}
\hat d^{ab}(k)=\frac{e^2}{3(2\pi)^3}\left(k^2g^{ab}-k^{a}k^{b}\right)\hat d_1(k)\ ,
\end{equation}
with $\hat d_1(k)$ the following distribution:
\begin{equation}\label{eq:8.5.13}
\hat d_1(k)=\left(1+\frac{2m^2}{k^2}\right)\text{sgn}\left(k_{-}\right)\Theta\left(k^2-4m^2\right)\sqrt{1-\frac{4m^2}{k^2}}\ .
\end{equation}
Now, a great amount of calculations could be avoided if one uses the following result, which can be shown by using the general splitting \textit{formulae} given in Eqs. \eqref{eq:3.6.9} and \eqref{eq:3.6.47}: \textit{For a causal distribution which in momentum space is of the form}
\begin{equation}\label{eq:8.2.100}
\hat d(p)=P(p)\hat d_1(p)\ ,
\end{equation}
\textit{with $P$ a polynomial, if $\hat r_1(p)$ is a retarded distribution corresponding to $\hat d_1(p)$, then $P(p)\hat r_1(p)$ is a retarded distribution corresponding to $\hat d(p)$}. This property is very convenient for practical purposes because it assures that it suffices to split the distribution $\hat d_1(p)$, which is less singular than $\hat d(p)$.

Therefore, returning to our problem, we only need to obtain the retarded part of $\hat d_1(k)$ given in Eq. \eqref{eq:8.5.13}, whose singular order at the $x^-$-axis is:
\begin{equation}\label{eq:8.5.14}
\omega_-\left[\hat d_1\right]=0\ .
\end{equation}
Its retarded part is given by --we use the variable $s=-2k_-q$--:
\begin{align}
\hat r_1&(k)=-\frac{i}{2\pi}\int\frac{ds}{s-ik_{-}0^+}\left\{\Theta(s+k^2-4m^2)\right.\nonumber\\
&\times\left(1+\frac{2m^2}{s+k^2}\right)\sqrt{1-\frac{4m^2}{s+k^2}}\nonumber\\
&\left.-\Theta(s-4m^2)\left(1+\frac{2m^2}{s}\right)\sqrt{1-\frac{4m^2}{s}}\right\}\ .\label{eq:8.5.15}
\end{align}
Applying Sokhotskiy's formula in the first integral, then changing the variable to $s+k^2\to s$ in it, we find:
\begin{align}
\hat r_1&(k)=\frac{i}{2\pi}k^2E(k)+\frac{1}{2}\text{sgn}\left(k_-\right)\Theta(k^2-4m^2)\nonumber\\
&\times\left(1+\frac{2m^2}{k^2}\right)\sqrt{1-\frac{4m^2}{k^2}}\ ,\label{eq:8.5.18}
\end{align}
with:
\begin{equation}\label{eq:8.5.17}
E(k)=\int\limits_{4m^2}^{+\infty}\frac{(s+2m^2)\sqrt{1-\dfrac{4m^2}{s}}}{s^2(k^2-s)}ds\ .
\end{equation}
This is the same integral which appears in instant dynamics \cite{ScharfFQED}, and has the value:
\begin{align}\label{eq:8.5.20}
E(k)=&\frac{m^2}{k^4}\left[\frac{1+\xi}{1-\xi}\left(\xi-4+\frac{1}{\xi}\right)\log(\xi)\right.\nonumber\\
&\left.+\frac{5}{3}\left(\xi+\frac{1}{\xi}\right)-\frac{22}{3}\right]\ ,
\end{align}
with the parameter $\xi$ defined by the relation:
\begin{equation}\label{eq:8.5.20-1}
\frac{k^2}{m^2}=-\frac{(1-\xi)^2}{\xi}\ .
\end{equation}
Therefore, the retarded distribution is:
\begin{align}
&\hat r_1(k)=\frac{i}{2\pi}\frac{m^2}{k^2}\left\{\left[\frac{1+\xi}{1-\xi}\left(\xi-4+\frac{1}{\xi}\right)\log(\xi)\right.\right.\nonumber\\
&\left.\left.+\frac{5}{3}\left(\xi+\frac{1}{\xi}\right)-\frac{22}{3}\right]-i\pi\text{sgn}\left(k_{-}\right)\right.\nonumber\\
&\left.\times\Theta(k^2-4m^2)(k^2+2m^2)\sqrt{1-\frac{4m^2}{k^2}}\right\}\ .\label{eq:8.5.21}
\end{align}
Note however that by Eq. \eqref{eq:8.5.20-1}:
\begin{equation}\label{eq:8.5.22}
\xi+\frac{1}{\xi}=2-\frac{k^2}{m^2}\ ,
\end{equation}
so the terms in the first line of Eq. \eqref{eq:8.5.21} which do not multiply the logarithm have coefficients subjected to normalization. Finally, putting this result into Eq. \eqref{eq:8.5.12} to obtain the retarded distribution $\hat r^{ab}(k)$ and subtracting the subsidiary distribution $\hat r'^{ab}(k)$, we are able to define the ``vacuum polarization tensor'' $\Pi^{ab}(k)$ as:
\begin{equation*}
\hat t^{ab}(k)=:-i\widehat\Pi^{ab}(k)\ ,
\end{equation*}
\begin{equation}\label{eq:8.5.23}
T_2^{(VP)}(x_1;x_2)=-i\normord{A_{a}(x_1)\Pi^{ab}(x_1-x_2)A_{b}(x_2)}\ ,
\end{equation}
so that:
\begin{equation}\label{eq:8.5.24}
\widehat\Pi^{ab}(k)=:(2\pi)^{-4}\left(\frac{k^{a}k^{b}}{k^2}-g^{ab}\right)\widehat\Pi(k)\ ,
\end{equation}
with:
\begin{align}
&\widehat\Pi(k)=\frac{e^2m^2}{3}\left\{\left[\frac{1+\xi}{1-\xi}\left(\xi-4+\frac{1}{\xi}\right)\log(\xi)\right.\right.\nonumber\\
&\left.+\frac{5}{3}\left(\xi+\frac{1}{\xi}\right)-\frac{22}{3}\right]\nonumber\\
&\left.-i\pi\Theta(k^2-4m^2)(k^2+2m^2)\sqrt{1-\frac{4m^2}{k^2}}\right\}\ .\label{eq:8.5.25}
\end{align}

Additionally, since $\widehat\Pi(k)$ has singular order $\omega_-=2$, its general expression is:
\begin{equation}\label{eq:8.5.26}
\widetilde\Pi(k)=\widehat\Pi(k)+C_0+C_2k^2\ ,
\end{equation}
because a term such as $c_{a}k^{a}$ is forbidden due to parity invariance of the QED. In order to fix the values of $C_0$ and $C_2$ we study M{\o}ller's scattering with vacuum polarization insertions. By a procedure identical to the one developed for the scattering of two fermions in Yukawa's model in Ref. \cite{AP2}, we find that the total radiation field propagator is the solution of the equation:
\begin{align}
\widehat D_{\text{tot}}^{ab}=\hat d^{ac}\left(\delta^b_c+(2\pi)^4\widetilde\Pi_{cd}\widetilde D_{\text{tot}}^{db}\right)\ ,\label{eq:8.5.27}
\end{align}
with $\hat d^{ab}$ the normalized distribution for M{\o}ller's scattering given in Eq. \eqref{eq:8.3.4}. Eq. \eqref{eq:8.5.27} can also be put in the following form:
\begin{equation}\label{eq:8.5.28}
\left(\delta^a_d-(2\pi)^4\hat d^{ac}\widetilde\Pi_{cd}\right)\widehat D_{\text{tot}}^{db}=\hat d^{ab}\ .
\end{equation}
The usual technique \cite{ScharfFQED} to solve this equation consists in inverting the distribution $\hat d^{ab}$. However, in our case this distribution has no inverse due to the non-covariant terms contained in it \footnote{This can be shown in the following way: Suppose that the inverse $\hat d^{-1}$ exists and write it as $\alpha g_{ab}+\beta k_bk_c+\gamma(\eta_b k_c+\eta_ck_b)+\delta\eta_b\eta_c$. The set of equations for the coefficients $\alpha,\ldots,\delta$ obtained from $\hat d^{-1}\hat d=1$ is inconsistent.}. Nonetheless, with Eqs. \eqref{eq:8.3.4} and \eqref{eq:8.5.24} we can form the inter-parenthetical expression of Eq. \eqref{eq:8.5.28}, which we will write as:
\begin{equation}\label{eq:8.5.29}
L\indices{^a_d}=\pi_1\delta^a_d+\pi_2k^a\eta_d\ ;
\end{equation}
\begin{equation*}
\pi_1=\frac{k^2-(2\pi)^{-2}\widetilde\Pi+i0^+}{k^2+i0^+}\ ,\ \pi_2=\frac{(2\pi)^{-2}\widetilde\Pi}{k_-(k^2+i0^+)}\ .
\end{equation*}
It turns that this tensor does have an inverse, which we will call $E\indices{^c_a}$:
\begin{equation}\label{eq:8.5.30}
E\indices{^c_a}=\sigma_1\delta^c_a+\sigma_2k^ck_a+\sigma_3k^c\eta_a+\sigma_4\eta^ck_a+\sigma_5\eta^c\eta_a\ .
\end{equation}
Then the coefficients $\sigma_i$ are found by the set of equations: $E\indices{^c_a}L\indices{^a_d}=\delta^c_d$; the solution is:
\begin{equation}\label{eq:8.5.31}
\sigma_1=\frac{1}{\pi_1}\ ,\ \sigma_2=0\ ,\ \sigma_3=-\frac{\sigma_1\pi_2}{\pi_1+k_-\pi_2}\ ,\ \sigma_4=0=\sigma_5\ .
\end{equation}
Substituting Eq. \eqref{eq:8.5.31} with the values of $\pi_i$ given in Eq. \eqref{eq:8.5.29} into Eq. \eqref{eq:8.5.30} we find:
\begin{equation}\label{eq:8.5.32}
E\indices{^c_a}=\frac{1}{k^2-(2\pi)^{-2}\widetilde\Pi+i0^+}\left\{k^2\delta^c_a-\frac{(2\pi)^{-2}\widetilde\Pi}{k_-}k^c\eta_a\right\}\ .
\end{equation}
Now we can solve Eq. \eqref{eq:8.5.28} by multiplying it by $E\indices{^c_a}$. We obtain that the total photon propagator is:
\begin{align}\label{eq:8.5.33}
\widehat{D}_{\text{tot}}^{cb}(k)=&-\frac{(2\pi)^{-2}}{k^2-(2\pi)^{-2}\widetilde\Pi(k)+i0^+}\nonumber\\
&\times\left(g^{cb}-\frac{k^c\eta^b+\eta^ck^b}{k_-}\right)\ .
\end{align}
As we can see, the total propagator preserves the same tensor structure of the distribution $\hat d^{ab}(k)$. This is different to what occurs when normalizing the total photon propagator in a covariant approach in instant dynamics, when the total propagator split into two terms: The one which contains $\widetilde\Pi$ is transversal to the momentum $k$, while the part parallel to the momentum remains independent of $\widetilde\Pi$ --see Ref. \cite{ScharfFQED}--. However, the two propagators reduce to the covariant one and are equal to each other once the conservation of the current is taking into account, eliminating all the terms proportional to $k^a$; this is an expression of gauge invariance.

The vacuum polarization scalar $\widetilde \Pi$ appears in the denominator of $\widehat{D}_{\text{tot}}^{ab}(k)$, so that it is possible to impose the physical requirements: (1) The physical mass of the photon is zero, so that the propagator must have a pole in $k^2=0$; (2) the physical value of the electric charge is the coupling constant $e$ of the one-point distribution $T_1$. These two requirements are translated, respectively, into:
\begin{equation}\label{eq:8.5.34}
\lim_{k^2\to 0}\widetilde\Pi(k)=0\quad\text{and}\quad\lim_{k^2\to0}\frac{d\widetilde\Pi(k)}{d(k^2)}=0\ .
\end{equation}
These two conditions are already satisfied by $\widehat\Pi(k)$ in Eq. \eqref{eq:8.5.25}, so that the coefficients in Eq. \eqref{eq:8.5.26} must be: $C_0=0=C_2$, and the right normalized solution is the central one.

\section{Conclusions}\label{sec:Conc}

We have formulated QED in light-front dynamics in the causal framework, for which we used the quantized field operators obtained by direct construction of Fock's space; it was proved that the equal-time (anti-) commutation relations for them are the same as the obtained in Refs. \cite{KogutSoper, Rohrlich2} and by the usage of Dirac-Bergmann's method and the correspondence principle in Ref. \cite{CPZ}.

We proved that M{\o}ller's and Compton's scattering processes are equivalent to those in instant dynamics if the right normalization terms are chosen, and, in the first case, if the conservation of current is taken into account --an extension off the mass-shell of the $S$ operator would lead to a difference with instant dynamics, but that is not manifest in the real world--. We can interpret this result by saying that the instantaneous terms in Feynman's propagators are not physical ones, but a consequence of the splitting procedure according to a time variable whose isochronic surfaces intersect the light-cone on the entire $x^-$-axis. Such a splitting procedure, by construction, cannot tell anything about the value of the retarded distribution at the $x^-$-axis \cite{APS22}, so the instantaneous terms that arise in it cannot be relied on, but must be fixed by other conditions besides causality. As we have seen, Lorentz's covariance implies that they must not be there. We see here that the intrinsic richness of the possible normalization terms in light-front dynamics allows one to start with an invariant $T_1$ distribution, without instantaneous interaction terms, that are unnecessary in order to obtain a covariant theory. They can be recovered, however, by defining the Lagrangian density as containing all the normalization terms of the higher-order transition distributions, which establishes a direct link to the usual approach, and showing in passing, and without the necessity of any combinatoric argument, that in a perturbation series based on $\mathscr L$, the instantaneous terms in it cancel exactly the ones in the field propagators.

In the study of vacuum polarization, the calculation is greatly simplified by the factorization of a second order polynomial, leading to a result which is equal to the one obtained in instant dynamics. For its normalization we have considered M{\o}ller's scattering with vacuum polarization insertions. This requires to define the total photon propagator, which has the same tensor structure as the commutation distribution of this field. Again, although different to the instant dynamics total propagator, it leads to the same physical results because the current conservation holds in the real world, as an expression of gauge invariance. The imposition of the zero mass of the photon and the value of the electric charge imply that the central solution is the right one.

Along this study we have encountered gauge invariance at two points: In the study of M{\o}ller's scattering and in the study of vacuum polarization. We have explicitly shown that the equivalence of these two results with instant dynamics relies on the gauge invariance property, expressed as the conservation of the electric current. Consequently, it is mandatory to study the complete implementation of quantum gauge invariance in null-plane QED. Our study of QED in the null-plane CPT will continue by addressing this problem and by considering other radiative corrections, Ward-Takahashi's identities, \textit{et cetera}.

\begin{acknowledgments}
O.A.A. thanks CAPES-Brazil for total financial support; B.M.P. thanks CNPq-Brazil for partial financial support.
\end{acknowledgments}

\bibliography{bibliographyAP}

\end{document}